\documentstyle[epsf,rotate]{EuroPhys}

\def\stars{\bigskip\centerline{***}\medskip}

\newif\ifboo \boofalse

\makeatletter
\def\@evenhead{{\normalsize\bf \thepage}\hfil}
\def\@maketitle{\rm\vbox to0pt{}\vskip-34pt
\parindent\z@
{\Large\bf \@title\par\smallskip}
\vskip 16pt \leftskip 25pt
{\normalsize\@author }
\vskip  4pt {\normalsize\it\@institute}
\vskip 16pt {\small\@pacs}
\vskip 36pt
\setcounter{page}{\value{startpage}}}
\makeatother
\Date{}
\shorttitle{H. SCHOMERUS: SEMICLASSICAL INTERFERENCE OF BIFURCATIONS}
\begin{document}

\title{Semiclassical interference of bifurcations}
\author{Henning Schomerus}
\institute{%
Fachbereich Physik,
	Universit\"at-Gesamthochschule Essen,
		D-45117 Essen, Germany}

\rec{}{}
\pacs{
\Pacs{05}{45.+b}{Theory and models of chaotic systems}
\Pacs{03}{65.Sq}{Semiclassical theories and applications}
\Pacs{03}{20.+i}{Classical mechanics of discrete systems: general
mathematical aspects}
}

\maketitle
\begin{abstract}
In semiclassical studies of 
systems with mixed phase space, the neighbourhood of 
bifurcations of co-dimension two is felt strongly even
though such bifurcations are ungeneric in classical mechanics.
We discuss a scenario which reveals this
fact
and derive the correct semiclassical contribution of the bifurcating orbits 
to the trace of the unitary time evolution operator.
That contribution has a certain collective character 
rather than being additive in the individual
periodic orbits involved.
The relevance of our observation
is demonstrated by a numerical study of the kicked top;
the collective contribution derived is found to considerably improve the
semiclassical approximation of the trace.
\end{abstract}


Periodic-orbit theory \cite{gutzwiller90}
aims at the semiclassical evaluation of energy
levels and relates the trace of the unitary time evolution operator
to periodic orbits of the corresponding classical systems.
A recent semiclassical study \cite{ktop93b}
based on periodic-orbit theory
was devoted to the neighbourhood, in the space of control parameters,
of classical bifurcations.
A collective treatment of the bifurcating orbits 
was found necessary, and even more the inclusion of 
predecessors of such orbits which live in complexified phase space and
were termed {\em ghosts}. The existing semiclassical studies
\cite{ktop93b,ozorio,sieber96,schomerussieber97}
focus on the generic
bifurcations in the classification of Meyer \cite{meyer70}.
The purpose of this Letter is to demonstrate that even the neighbourhood
of a bifurcation which is ungeneric to the classical system
is felt semiclassically and necessitates a collective treatment.
This implies that
collective contributions of this kind will constitute a basic ingredient
in a semiclassical trace formula for systems with mixed phase space.

The situation we have in mind manifests itself in certain sequences of
bifurcations. Sadovski{\'\i}, Shaw, and Delos \cite{Sadovskii95}
found that such sequences can be explained
by normal-form theory \cite{poincare} and
mentioned the importance of their observation for semiclassical
studies of systems with mixed phase space. We go one step further and
derive the explicit collective 
semiclassical amplitude of a group of orbits involved
in the bifurcations. 

We shall 
concentrate on a certain sequence of bifurcations,
that of a period tripling and the tangent bifurcation of the
satellite. In the diamagnetic Kepler problem considered in \cite{Sadovskii95}
this scenario was not found. 

The semiclassical expression to be established will be tested for 
the periodically kicked top \cite{ktop}.
A period-to-period stroboscopic description of the quantum mechanics
involves the unitary Floquet operator
\[
F=\exp(-i\frac{k_z}{2j+1}J_z^2-ip_zJ_z)
\exp(-ip_yJ_y)
\exp(-i\frac{k_x}{2j+1}J_x^2-ip_xJ_x)\;.
\]
We here encounter angular-momentum operators $J_{x,y,z}$ obeying the
commutator relation
$[J_k,J_l]=i\epsilon_{ijk}J_k$. The Hilbert space dimension is fixed
as $2j+1$ (with $j$ the good quantum number from ${\bf J}^2=j(j+1)$),
and the semiclassical limit is reached by
sending $j\to\infty$.
The classical phase space is the sphere ${\bf J}^2/(j(j+1))=1$.
The parameters $p_i$ may be interpreted as angles of rotation
while the $k_i$ characterize nonlinear rotations sometimes called
torsions. To steer the system towards 
bifurcations, we hold the $p_i$ at
fixed values and vary $k=k_z=10 k_x$;
for $k=0$ the system is integrable while for $k=5$ it
displays well developed chaos.
Neither time reversal nor any geometrical symmetry is present.

\begin{figure}
\epsfysize14cm
\rotate[r]{\epsfbox{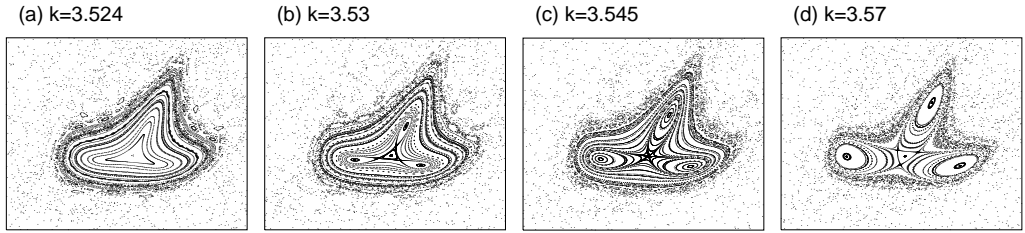}}
\caption{Section of the phase space displaying the sequence of
bifurcations. In (a) a stable orbit is surrounded by its stability
island.
From (a) to (b) 
two satellites appear via a tangent bifurcation, the inner one
approaching
the central orbit in the period tripling (c) and finally
re-emerging on the opposite side (d).}
\label{phasespace}
\end{figure}

The situation for the period tripling
is visualized in the sequence of phase-space portraits 
in fig.\ \ref{phasespace}.
Initially (a), 
a stable periodic orbit of period one
is surrounded by its stability
island. At a certain value of the control parameter ($k^{(\rm tan)}=3.525$)
two satellites of triple period come into existence via a tangent
bifurcation. Then the inner (unstable) satellite approaches the
central orbit (b), collides with it in the period tripling at $k^{(\rm trip)}=3.544$ (c), and
finally re-emerges on the other side (d).
Astonishing closeness
of the bifurcations in parameter space seems not exceptional even
in the regime of well developed chaos: Starting with $k=0$
we found the
first four period triplings of orbits with primitive period one
all accompanied by a tangent bifurcation at the pairs
($k^{(\rm tan)}=1.853$, $k^{(\rm trip)}=1.859$), 
($k^{(\rm tan)}=1.971$, $k^{(\rm trip)}=1.975$),
the one mentioned above,
and
($k^{(\rm tan)}=9.334$, $k^{(\rm trip)}=9.370$).
It has to be stressed that {\em no second parameter was tuned}  
in the present study to achieve such closeness.
However, if one
varied a second control parameter,
the system could be tuned
such that both bifurcations happen at the same point
in  parameter space. This means that
one deals with the {\em unfolding of a bifurcation of co-dimension two}.

The classical stroboscopic map $(q,p)\to(q^\prime,p^\prime)$  (for one degree of freedom)
has the classical action
$S(q^\prime,p)$ as the generating function, with
$S_{q^\prime}=p^\prime$, $S_p=q$ (here and in the following we denote
partial derivatives by indices).
Close to the bifurcations ($k\approx k^{\rm (tan,\,trip)}$) and in the neighbourhood of the
bifurcating orbits, the action can be approximated by
normal forms \cite{poincare}.
They contain the relevant information
about the configuration in phase space
and the scaling properties of the classical quantities.
For the period tripling (without the tangent bifurcation of the
satellites) one can use action-angle coordinates $I, \varphi$ with
$q=\sqrt{2I}\cos\varphi$ and
$p=\sqrt{2I}\sin\varphi$
and has
\begin{equation}
\label{normalform}
S(I,\varphi^\prime)=I\varphi^\prime+\epsilon I +a I^{3/2} \cos 3\varphi^\prime\;.
\end{equation}
Here $\epsilon=k-k^{(\rm trip)}$ is the control parameter; at $\epsilon=0$ the period tripling takes
place. The coordinates of the satellite periodic orbits are given by
$S_I=\varphi^\prime$, $S_{\varphi^\prime}=I$;
hence, the $\varphi$-coordinate obeys
$
-3a I^{3/2} \sin 3\varphi^\prime =0
$.
Since a threefold symmetry is implied by this equation 
it suffices 
to consider the second equation $S_I=0$ on the $q$-axis
after switching back to the coordinates $p,q$, yielding
$ \epsilon q+\frac 3 {\sqrt 8} a q^2=0 $.
This describes the
central orbit at $I=0$ and the unstable satellite at distance $I=q^2/2=4\epsilon^2/(9a^2)$.

The normal form
originally stems from a Taylor--Fourier expansion of the full action
around the central orbit, followed  by a certain rectification 
procedure. While this series generally does not converge
one still can attempt to include higher order terms to explain certain
phenomena (see also \cite{sieber96,schomerussieber97}), as was argued
for the inclusion of further satellites by the authors of
\cite{Sadovskii95}.

The next-order term is $b I^2$, and the
extended normal form is
\begin{equation}\label{extendednform}
S(I,\varphi^\prime)=I\varphi^\prime+\epsilon I +a I^{3/2} \cos 3\varphi^\prime+bI^2\;.
\end{equation}
The $\varphi$-coordinate of the periodic orbits once more obeys
$ -3a I^{3/2} \sin 3\varphi^\prime =0 $.
On the $q$-axis they now satisfy
$ \epsilon q+\frac 3{\sqrt 8} a q^2 + b q^3=0 $.
This equation has three solutions,
\[
q_0=0\;, \quad q_{\pm}=-\frac {3}{4\sqrt2} \frac a b \pm \sqrt{\frac 9
{32} \frac {a^2}{b^2}-\frac {\epsilon}{b}}\;.
\]
One in fact sees that the inclusion of the next-order term 
implies the existence of a further satellite. 
We will denote the central orbit by the index $0$ and the satellites
by $\pm$ as it is done here for the coordinates.
At $\epsilon^{(\rm tan)}=9a^2/(32b)$ the satellites undergo a tangent
bifurcation and for $\epsilon/b>9a^2/(32b^2)$ both satellites are ghosts (cf.
Fig.\ \ref{phasespace} (a)).
For $0<\epsilon/b<9a^2/(32b^2)$ both satellites are on the same side of the 
central orbit (Fig.\ \ref{phasespace} (b)), while after the period tripling ($\epsilon=0$)
 they lie opposite to each other, Fig.\ \ref{phasespace} (d).
The situation approaches that of a broken torus as $\epsilon/b\to -\infty$.
When a second parameter is varied to
achieve $a=\epsilon=0$ both satellites are contracted onto the central orbit
in a co-dimension two bifurcation.

Our semiclassics for maps starts with the expression
\cite{ozorio,berrymount72,junkerleschke92}
\begin{equation}
\label{scstart}
\mbox{tr}\, F^n=\int\!\!\int\frac{{\rm d} q^\prime\,{\rm d} p}{2\pi\hbar}
|S^{(n)}_{q^\prime p}|^{1/2} \exp\left\{\frac i \hbar (S^{(n)}-q^\prime p)- i \frac
\pi 2 \nu\right\}
\end{equation}
of the trace of the Floquet operator $F$ as an integral over
phase space.
It involves the classical action $S^{(n)}(q^\prime,p)$ generating the 
$n$-th iteration of the
classical map
and a topological phase expressed by the Morse index $\nu$.
The orbits give a semiclassical contribution 
$C^{(\rm cluster)}$ to $\mbox{tr}\,F^n$ if $n$ is a multiple of their
primitive periods $n^{(0)}$, related to each other by
$n^{(0)}_+=n^{(0)}_-= 3 n^{(0)}_0$.

The asymptotic behaviour for large $\epsilon/\hbar$ 
is found with the stationary-phase approximation (spa)
\cite{gutzwiller90,junkerleschke92},
\begin{equation}
\label{asympt}
C^{(\rm cluster)}_{\rm spa} = \!\!\!\sum_{k=\{0,\pm\}}\!\!
\frac{n^{(0)}_k}{|2-\mbox{tr}\, M_k|^{1/2}}\exp\left\{\frac i \hbar
S_k-i\frac\pi 2\mu_k\right\}\;,
\end{equation}
a sum of three individual
contributions, one for each orbit.
The $2\times2$ stability matrix $M$ describes the
linearized map at the locus of the orbit in phase space.
The Maslov indices are $\mu_0=\nu-\mbox{sign}\,\epsilon$,
$\mu_\sigma=\nu$,
$\mu_{-\sigma}=\nu-\mbox{sign}\,b$ with $\sigma=\mbox{sign}\,ab$.

Introducing in (\ref{scstart}) the normal form (\ref{extendednform})
into the exponent with
$\sqrt{S_{I,\varphi^\prime}}\approx 1$ 
one obtains the {\em collective semiclassical contribution  $C^{(\rm cluster)}_{\rm local}$} of the
bifurcating orbits to $\mbox{tr}\,F^n$ in the {\em local approximation}
valid
close to the bifurcations ($\epsilon/\hbar$
small). The coefficients in the normal form have to be expressed
by the actions $S_k$ of the orbits in order to obtain 
a contribution that is invariant under canonical transformations
\cite{sieber96,schomerussieber97,tomsovic95}. 

One can find a uniform approximation which interpolates between the local
 and the stationary-phase approximation
by expanding
$\sqrt{S_{I,\varphi^\prime}}\approx 1+ BI + CI^2$ where the corrections come 
from the higher order terms of the normal form.
The uniform approximation found is
\begin{equation}\label{uniform}
C^{(\rm cluster)}_{\rm uniform}=\int\!\!\int
\frac{{\rm d}\varphi^\prime\,{\rm d}I}{2\pi\hbar}
(1+BI+CI^2)
\exp\left\{\frac i \hbar(\epsilon I+aI^{3/2}\cos 3\varphi^\prime+bI^2)- i \frac
\pi 2 \nu\right\}\;
\end{equation}
and
is valid for arbitrary combinations of $\epsilon$ and $\hbar$ as
these parameters approach zero.
The coefficients $B$ and $C$ have to be
determined such that (\ref{asympt}) is 
recovered as $\hbar\to 0$.

In the limit $\epsilon /b\to-\infty$
the satellites form a broken torus, well separated from the central orbit.
To discuss this case it is useful
to cast the prefactor into the alternative
form $1+B^\prime I+C^\prime I^{3/2}\sin 3\varphi^\prime$
by partial integration. One then recovers
the uniform approximation found by Tomsovic, Grindberg, and Ullmo \cite{tomsovic95}.

Let us sketch how one arrives at a form of the integral 
(\ref{uniform}) which makes
a numerical evaluation tractable.
The integration over $\varphi^\prime$ gives
a Bessel function, and after proper rescaling of $I$ one arrives at an
integral of type
$
K(\epsilon ,b)=
\int_0^\infty {\bf J}_0(I^{3/2}) \exp[i(\epsilon  I+ bI^2)]\,{\rm d} I
$
for the local approximation. 
Expanding the exponential function and with the help of
identities for the $\Gamma$-function at double and triple argument
\cite{gradshteyn} we obtain a useful representation in terms of a double sum,
\begin{eqnarray}\label{series}
&&K(\epsilon ,b)=
\frac{\sqrt\pi}{2}\sum_{n=0}^\infty\sum_{m=0}^\infty
\frac{(i\epsilon )^n}{n!m!}(-ib)^{-(n+1)/2}(-4ib)^{-3m}
\\
&&\quad
\times
\left(\frac{\Gamma\left(\frac{n+1}{2}+3m\right)}{
(2m)!\Gamma\left(\frac1 2 +m\right)}
-\frac{(4b/i)^{-3/2}\Gamma\left(\frac n 2 + 2 + 3m\right)}{
(2m+1)!\Gamma\left(\frac 3 2 + m\right)}
\right)\;.
\nonumber
\end{eqnarray}
The two additional integrals 
in the uniform approximation can be found by taking the derivative with
respect to $i\epsilon $.

\begin{figure}
\epsfysize7cm
\rotate[r]{\epsfbox{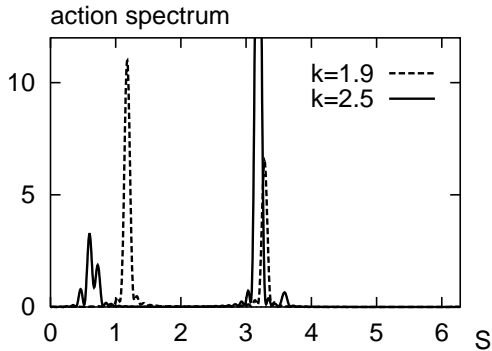}}
\caption{The action spectrum $|T^{(3)}(S)|^2$
as the systems is steered across bifurcations.
The peaks at $S\approx 1$ belong to the three orbits which take part
in the bifurcations at $k^{(\rm tan,\,trip)}\approx 1.85$,
the other peak to another cluster of
orbits bifurcating at $k^{(\rm tan,\,trip)}\approx 1.97$.}
\label{ftrafo}
\end{figure}

\begin{figure}
\epsfysize14cm
\rotate[r]{\epsfbox{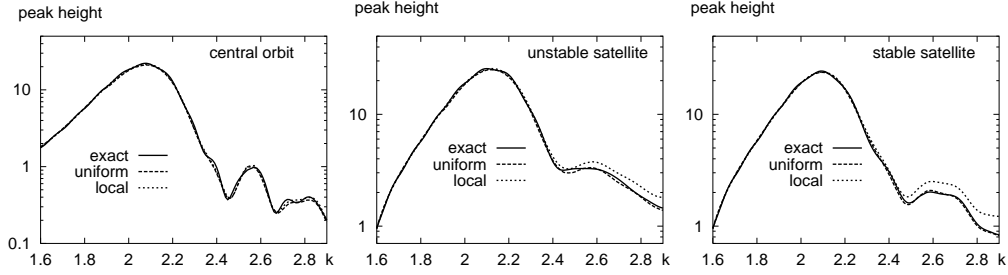}}
\caption{Quantum mechanically exact and semiclassically approximated
peak heights in the action spectrum
at the values of the classical actions of the three 
orbits which bifurcate at $k\approx 1.85$,
plotted as a function of the control parameter $k$.
Both the uniform and the local approximation
work well close to the bifurcations;
the uniform approximation remains valid even far away from the
bifurcations.}
\label{peaks1}
\end{figure}

To investigate the quality of the various semiclassical approximations
we now turn to the study of the
{\em action spectrum}, that is the Fourier expansion of $\mbox{tr}\,F^n$
with respect to
$j\equiv 1/\hbar$. Specifically, we study the function \cite{actionspectrum}
\[
T^{(n)}(S)=\frac 1 {64} \sum_{j=1}^{64}e^{-ijS}\mbox{tr}\, F^n(j)\;.
\]
The squared modulus of this
function shows peaks at values of $S$ corresponding to
values of the classical actions of periodic orbits, as can be seen 
using the asymptotic behaviour (\ref{asympt}) of the contribution
$C^{(\rm cluster)}$ to $\mbox{tr}\, F^n$.
Fig.\ \ref{ftrafo} illustrates this for the quantum mechanical exact $|T^{(3)}(S)|^2$
and values of the control parameter close to the period
tripling at $k^{(\rm trip)}=1.859$.
The peak at $S\approx 1$ corresponds to the bifurcating orbits.
If one steers away from 
the bifurcations
the peak is resolved into three distinct ones at the classical actions
of the bifurcating orbits.
It is instructive to calculate the height of the peaks
both quantum-mechanically exact as well as
semiclassically as $k$
is varied across the bifurcations.
Fig.\
\ref{peaks1} displays the result obtained with the uniform and the local
approximation. 
Close to and on the left of the bifurcations
both collective contributions are in such good
coincidence with the exact result
that the difference is
hard to resolve.
Far away from the bifurcations the local approximation starts to
fail while the uniform approximation remains valid.

\begin{figure}
\epsfysize7cm
\rotate[r]{\epsfbox{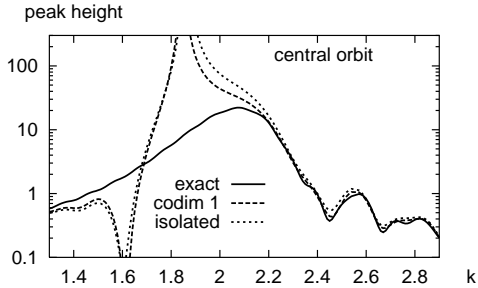}}
\caption{Peak height of the central orbit
as in fig.\ \protect\ref{peaks1}. Both the
co-dimension-one approximation
as well as the approximation that treats all orbits as
isolated 
only work well far away from the bifurcations.}
\label{peaks2}
\end{figure}

Fig.\ \ref{peaks2} reveals that the three orbits
necessarily have to be treated together close to
the bifurcations.
For $k>k^{(\rm trip)}=1.859$, the co-dimension-one approximation treats 
the
outer, stable satellite as isolated
and the central orbit collectively together with the unstable
satellite by a uniform approximation discussed in
\cite{schomerussieber97}.
For $k<k^{(\rm tan)}=1.853$, this approximation
considers the central orbit as isolated
and incorporates the satellites collectively
\cite{ktop93b,schomerussieber97,ozorio}.
It fails close to the bifurcations and only regains
validity far away.
There the orbits can also
be treated individually and
the stationary-phase approximation becomes justified.

From the comparison of figs.\ \ref{peaks1} and \ref{peaks2} it is apparent that
the interval where all three orbits have to be treated collectively 
is quite large. 
It follows
that unfoldings of bifurcations of higher co-dimension are
relevant for semiclassical studies even if one varies only a single control
parameter. 
This can be understood recalling that
there exists
a semiclassical measure of vicinity of
orbits, the difference of their actions in units of Planck's constant,
$\Delta S/(2\pi\hbar)$. This measure is unknown to classical mechanics
such that there bifurcations at different parameter values  do not feel each
other so as to necessitate a qualitatively different treatment.

In conclusion, in this Letter we studied
configurations of periodic orbits
that participate in sequences of bifurcations. It
was shown that in semiclassical studies involving periodic-orbit theory
such configurations have to be treated collectively.
This was illustrated for the sequence of a
period tripling and the tangent bifurcation of the satellite.
A uniform collective contribution was given and tested for the kicked top.
We found excellent agreement between semiclassical and exact peak
heights in the action spectrum.

In the language of normal-form theory, the case studied here
is the unfolding of a period-three bifurcation of
co-dimension two.
Normal-form theory predicts similar scenarios for every period-$n$ bifurcation,
and indeed further examples are under investigation.

\stars

The author has the great pleasure to thank Fritz Haake, Martin Sieber,
G{\"u}nter Wunner,
and Marek Ku\'s for enlightening discussions.
Support by the Sonderforschungsbereich
`Unordnung und gro{\ss}e Fluktuationen' of the Deutsche Forschungsgemeinschaft
is gratefully acknowledged.



\end{document}